\documentclass[12pt]{article}
\usepackage{enumerate}
\usepackage{ifpdf}
\usepackage{ae}
\usepackage[T1]{fontenc}
\usepackage[ansinew]{inputenc}
\usepackage{amsmath}
\usepackage{relsize}
\usepackage{amssymb}
\usepackage{tikz}
\usepackage{graphicx}
\usepackage{color}
\definecolor{darkblue}{cmyk}{0.9,0.9,0,0}
\definecolor{darkgreen}{rgb}{0,0.55,0}
\usepackage[colorlinks=true,linkcolor=darkblue,citecolor=darkblue,urlcolor=darkblue]{hyperref}
\usepackage{epsfig}
\usepackage{epstopdf}
\usepackage{graphicx}

\makeatletter
\long\def\@makecaption#1#2{
  \vskip\abovecaptionskip
  \sbox\@tempboxa{{\captionfonts #1: #2}}
  \ifdim \wd\@tempboxa >\hsize
    {\captionfonts #1: #2\par}
  \else
    \hbox to\hsize{\hfil\box\@tempboxa\hfil}
  \fi
  \vskip\belowcaptionskip}
\makeatother

\newcommand{\beq}{\begin{equation}}
\newcommand{\eeq}{\end{equation}}
\newcommand{\beqy} {\begin{eqnarray}}
\newcommand{\eeqy} {\end{eqnarray}}
\newcommand{\bsmat}{\begin{smallmatrix}}
\newcommand{\esmat}{\end{smallmatrix}}
\newcommand{\bmat}{\begin{matrix}}
\newcommand{\emat}{\end{matrix}}

\def\({\left(}
\def\){\right)}
\def\[{\left[}
\def\]{\right]}

\def\<{\langle}
\def\>{\rangle}

\usepackage{varioref}
\usepackage{makeidx}
\makeindex

\usepackage[english]{babel}
\usepackage{caption}
\usepackage{subfig}

\usepackage{tabularx}
\begin{document}

\thispagestyle{empty}

\renewcommand{\thefootnote}{\fnsymbol{footnote}}
\setcounter{page}{1}
\setcounter{footnote}{0}
\setcounter{figure}{0}
\begin{center}
$$$$

{\LARGE\textbf{\mathversion{bold}
The superconformal bootstrap \\ for structure constants}}
\vspace{1.0cm}

\textrm{\Large Luis F. Alday and Agnese Bissi}
\\ \vspace{1.2cm}

\textit{Mathematical Institute, University of Oxford,}  \\
\textit{Radcliffe Observatory Quarter, Oxford, OX2 6GG, UK} \\
\vspace{5mm}

\par\vspace{1.5cm}

\textbf{Abstract}\vspace{2mm}
\end{center}

\noindent
We report on non-perturbative bounds for structure constants on ${\cal N}=4$ SYM. Such bounds are obtained by applying the conformal bootstrap recently extended to superconformal theories. We compare our results with interpolating functions suitably restricted by the S-duality of the theory. Within numerical errors, these interpolations support the conjecture that the bounds found in this paper are saturated at duality invariant values of the coupling. This extends recent conjectures for the anomalous dimension of leading twist operators.

\vspace*{\fill}

\setcounter{page}{1}
\renewcommand{\thefootnote}{\arabic{footnote}}
\setcounter{footnote}{0}

\newpage

 \def\nref#1{{(\ref{#1})}}



\section{Introduction: the superconformal bootstrap}

{\bf The conformal bootstrap}

\noindent The most natural observables in a conformal field theory (CFT) are correlators of gauge invariant local operators. The operator product expansion (OPE) can be used in order to compute any $n-$point correlation function in terms of the CFT data: the spectrum of anomalous dimensions and the structure constants. The basic idea of the conformal bootstrap is to constrain the CFT data by using symmetries of correlation functions, together with unitarity and the structure of the OPE. For instance, let us consider the four-point function of a scalar field $\phi$ of dimension $d$ in a generic CFT. Conformal symmetry implies
\begin{equation}
\langle \phi(x_1) \phi(x_2) \phi(x_3) \phi(x_4) \rangle = \frac{g(u,v)}{x_{12}^{2d} x_{34}^{2d}}
\end{equation}
where we have introduced the cross-ratios $u=(x_{12}^2 x_{34}^2)/(x_{13}^2x_{24}^2)$ and $v=(x_{14}^2 x_{23}^2)/(x_{13}^2x_{24}^2)$. By considering the OPE $\phi(x_1) \times \phi(x_2)$ we can decompose the four point function into conformal blocks
\begin{equation}
\label{opedec}
g(u,v) = 1 +\sum_{\ell,\Delta} a_{\Delta,\ell} g_{\Delta,\ell}(u,v)
\end{equation}
the sum runs over the tower of primaries present in the OPE ( ${\cal O}_{\Delta,\ell} \in \phi \times \phi$ ) and $\ell$ and $\Delta$ denote the spin and the dimension of the intermediate primary. $a_{\Delta,\ell}=c_{\Delta,\ell}^2$ denotes the square of the structure constants and is non-negative due to unitarity. The conformal blocks $g_{\Delta,\ell}(u,v)$ are explicitly known functions, fixed by conformal symmetry, which repack the contribution of all descendants of a given primary. Finally, we have singled out the contribution from the identity operator.

We could have instead considered the OPE $\phi(x_2) \times \phi(x_3)$. Crossing-symmetry of the four-point function 
\begin{equation}
 \frac{g(u,v)}{x_{12}^{2d} x_{34}^{2d}}= \frac{g(v,u)}{x_{23}^{2d} x_{14}^{2d}}~ \rightarrow ~v^d g(u,v) = u^d g(v,u)
\end{equation}
together with associativity of the OPE imply the conformal bootstrap equation
\begin{eqnarray}
&\sum_{\ell,\Delta} a_{\Delta,\ell} F_{\Delta,\ell}(u,v) =1,~~~~~~a_{\Delta,\ell}  \geq 0\\
&F_{\Delta,\ell}(u,v) \equiv  \frac{v^d g_{\Delta,\ell}(u,v)- u^d g_{\Delta,\ell}(v,u)}{u^d-v^d} \nonumber
\end{eqnarray}
In $\cite{Rattazzi:2008pe}$ it was understood how to use efficiently this equation in order to find non-perturbative bounds for the anomalous dimensions of operators appearing in the OPE $\phi \times \phi$. The idea is the following. A given trial spectrum of dimensions and spins $\{\Delta,\ell\}$ can be ruled out if we can find a linear operator $\Phi$ such that
\begin{eqnarray}
\Phi(F_{\Delta,\ell}(u,v)) &\geq& 0,~~~~\mbox{for}~a_{\Delta,\ell} \neq 0\\
\Phi(1) &<& 0 \nonumber
\end{eqnarray}
since this would mean the bootstrap equation has no solutions with $a_{\Delta,\ell}$ non-negative. By considering families of trial spectra it is possible to put bounds on the dimensions of the leading twist operator for a given spin $\ell$. A similar idea can be used to put bounds on the OPE coefficients\footnote{Sometimes we will call OPE coefficient to $a_{\Delta,\ell}$ with the understanding that it is the square of the OPE coefficient.} $a_{\Delta,\ell}$, see {\it e.g.} \cite{Poland:2010wg,ElShowk:2012hu}. We single out a particular dimension and spin $\{\Delta_0,\ell_0\}$ and then normalize the linear operator such that $\Phi(F_{\Delta_0,\ell_0}(u,v))=1$. If $\Phi(F_{\Delta,\ell}(u,v)) \geq 0$ for all other values of $\ell$ and $\Delta$, we obtain the following bound
\begin{equation}
a_{\Delta,\ell} = \Phi(1) - \sum_{\ell,\Delta \neq \ell_0,\Delta_0} a_{\Delta,\ell} \Phi(F_{\Delta,\ell}(u,v)) \leq  \Phi(1)
\end{equation}
Note that for every "allowed" spectrum $\Phi(1)$ will be positive. 

The conformal bootstrap has been used successfully in order to constraint the CFT data for several CFT's in various dimensions. These constraints can be very powerful and in some cases they can be even enough to fix it completely! see {\it e.g.} \cite{Rychkov:2009ij, ElShowk:2012ht}.

Over the last few years, there has been remarkable progress in the computation of observables in ${\cal N}=4$ super-Yang-Mills (SYM). It has also become apparent that the theory has a tremendously rich structure. Much of this progress, however, has been confined to either perturbation theory (at weak or at strong coupling) in the planar limit, or to protected quantities. In  \cite{Beem:2013qxa} it was shown how the methods of the conformal bootstrap can be applied to ${\cal N}=4$ SYM.

\bigskip

\noindent {\bf The superconformal bootstrap}

\noindent The starting point is the four-point correlator of the protected scalar operator transforming in the ${\bf 20}'$ of the $SU(4)$ R-symmetry group ${\cal O}^{(IJ)}_{{\bf 20}'}=Tr \phi^{(I} \phi^{J)}$, $I=1,...,6$. This correlator has been studied in much detail, see for instance \cite{Eden:2000bk,Dolan:2001tt}. The OPE ${\cal O}^{(IJ)}_{{\bf 20}'} \times {\cal O}^{(KL)}_{{\bf 20}'}$, and hence the four point function, receives contributions from long multiplets, as well as from short and semi-short multiplets. It turns out that crossing-symmetry plus superconformal symmetry can be used to fix the contributions from short and semi-short multiplets  \cite{Beem:2013qxa}. Furthermore, for long multiplets, the superconformal primary is always a $SU(4)$ singlet \cite{Eden:2001ec,Dolan:2001tt}. The conformal bootstrap equation for ${\cal N}=4$ SYM takes the final form \cite{Beem:2013qxa}:
\begin{equation}
\label{super}
\sum_{\substack{\ell=0,2,..., \\ \Delta \geq \ell+2}} a_{\Delta,\ell} F_{\Delta, \ell}(u,v)=F^{short}(u,v,c)
\end{equation}
$a_{\Delta,\ell}$ denotes the (square of the) structure constant involving two protected operators  ${\cal O}_{{\bf 20}'}$ and a superconformal primary, singlet of $SU(4)$, of spin $\ell$ and dimension $\Delta$. $F^{short}(u,v,c)$ is an explicit, calculable function, arising from short and semi-short contributions and is the analogue of the $1$ in the right hand side of the usual conformal boostrap equation \footnote{We thank L. Rastelli for sharing with us their closed expression for $F^{short}$.}. It is independent of the coupling constant and depends on the gauge group only through its central charge $c$. For $SU(N)$ gauge groups we have $c =(N^2-1)/4$. Finally, we have defined
\begin{equation}
\begin{aligned}
F_{\Delta,\ell}(u,v) = v^2 u^{\frac{\Delta-\ell}{2}} G_{\Delta+4}^{(\ell)}(u,v)-u^2 v^{\frac{\Delta-\ell}{2}} G_{\Delta+4}^{(\ell)}(v,u)\\
G_{\Delta}^{(\ell)} = \frac{1}{z-\bar z}\left( (-\frac{z}{2})^\ell z k_{\Delta+\ell}(z) k_{\Delta-\ell-2}(\bar z) - (z \leftrightarrow \bar z) \right)
\end{aligned}
\end{equation}
with $k_\beta(z) = ~_2F_1(\beta/2,\beta/2,\beta;z)$ and we have used $u =z \bar z,~~v=(1-z)(1-\bar z)$.

In \cite{Beem:2013qxa} the superconformal bootstrap equation (\ref{super}) was used in order to obtain numerical bounds for the anomalous dimensions of leading twist operators of low spin. Those bounds provided true non-perturbative information about non-planar ${\cal N}=4$ SYM. In this note, we report a modest generalization of their results. We use the superconformal bootstrap equation (\ref{super}) in order to find numerical bounds to structure constants of  ${\cal N}=4$ SYM.

In the next section we derive global/exact bounds, valid for any value of the coupling constant and making no assumptions about the spectrum of the theory. In section three, we focus in two special cases where something can be said about the spectrum: the free theory and S-duality invariant points. At the end of section three we compare our results with interpolating functions constructed from the available perturbative data. Finally we end up with some discussion. In the appendices we include some technical details as well as numerical results for the bounds in several cases. 

\section{Global bounds}

We start by considering eq. (\ref{super}) at generic values of the coupling constant. In this case, $\ell=0,2,4,...$ but $\Delta$ can take any continuos values, provided $\Delta\geq \ell+2$.  In order to write down our linear operator, it is convenient to use variables $a,b$ such that 
$$z= 1/2+a+b,~~~~~\bar{z} = 1/2+a-b$$
Due to the symmetries of the correlator, both $F_{\Delta,\ell}$ and $F^{short}$ are odd functions of $a$ and even functions of $b$. The linear operators we consider take the following form
\begin{equation}
\Phi^{(\Lambda)} f(a,b) \equiv \sum_{i,j=0}^{2i+2j+1=\Lambda} \frac{ \kappa_{ij}}{(2i+1)!(2j)!}\partial_a^{2i+1}\partial_b^{2j}f(a,b)|_{a=b=0}
\end{equation}
Namely, we expand around the symmetric point $z=\bar z=1/2$. This is an obvious choice, but most importantly, one can check numerically, for specific cases, that acting with this operator and summing over $\ell$ and $\Delta$ in (\ref{super}) commute. In particular, after acting with $\Phi^{(\Lambda)}$ on each $F_{\Delta,\ell}$ , we expect the sum over $\ell,\Delta$ to converge. 

Once we have selected a particular $\{\Delta_0,\ell_0\}$, the task is then to minimize $\Phi^{(\Lambda)} F^{short}$ subject to the constraints $\Phi^{(\Lambda)} F_{\Delta_0,\ell_0} = 1$ and $\Phi^{(\Lambda)} F_{\Delta,\ell} \geq 0$ for all $\{\Delta,\ell \} \neq \{\Delta_0,\ell_0\}$.
In practice we use cut-offs  $\ell =0,2,..., \ell_{max},~~~0 \leq \Delta-\ell-2 \leq \Delta_{max}$ and discretize $\Delta$. For the results shown in this note we have used $\ell_{max} = \Delta_{max} = 20$ and $\delta \Delta = 1/8$. We have explicitly checked that changing theese cut-offs does not improve the bounds significantly. Furthermore, the constraints inside the cut-off are supplemented by the asymptotic constraints. Assuming $\ell,\Delta \gg 1$ with $\ell = \alpha \Delta$, we find 

\begin{equation}
\Phi^{(\Lambda)} F_{\Delta,\ell=\alpha \Delta} \sim \sum_{2i+2j+1=\Lambda} \frac{(\Lambda+1)!}{(2i)!(2j+2)!} \alpha^{2j+2} \kappa_{ij} \equiv \Phi^{(\Lambda)}_{asymp}(\alpha)
\end{equation} 
The asymptotic constraints take the form
\begin{equation}
\label{asymp}
\Phi^{(\Lambda)}_{asymp}(\alpha) \geq 0~~~~\mbox{for}~0\leq \alpha \leq 1
\end{equation}
For a fixed spin $\ell_0$ we assume that the leading twist operator has dimension $\Delta_0$, with $\ell+2 \leq \Delta_0 \leq \Delta_{l}^c$, where $\Delta_{l}^c$ is the bound for the dimension . Then we need to solve the following optimization problem

\begin{eqnarray*}
\mbox{Minimize~} ~~~&&\Phi^{(\Lambda)} F^{short} \\
\mbox{subject to} ~~~ &&\Phi^{(\Lambda)} F_{\Delta_0,\ell_0} = 1, \\
&&\Phi^{(\Lambda)} F_{\Delta,\ell_0} \geq 0,~~~~~ \Delta > \Delta_0\\
&&\Phi^{(\Lambda)} F_{\Delta,\ell}  \geq 0,~~~~~~ \ell \neq \ell_0,~ \Delta \geq \ell+2\\
&&\Phi^{(\Lambda)}_{asymp}(\alpha) \geq 0,~~~~\mbox{for all}~0\leq \alpha \leq 1
\end{eqnarray*}
This optimization problem involves $\frac{1}{8}(\Lambda+1)(\Lambda+3)$ variables $\kappa_{ij}$ and can be solved numerically using standard linear programming software. We have worked with $\Lambda=17$. Fig. 1 shows the exclusion regions for $a_{\Delta,\ell}$ as a function of $\Delta$, for $\ell=0,2$ and different values of the central charge.  

\begin{figure}[h]
\begin{center}$
\begin{array}{ccc}
\includegraphics[width=3.25in]{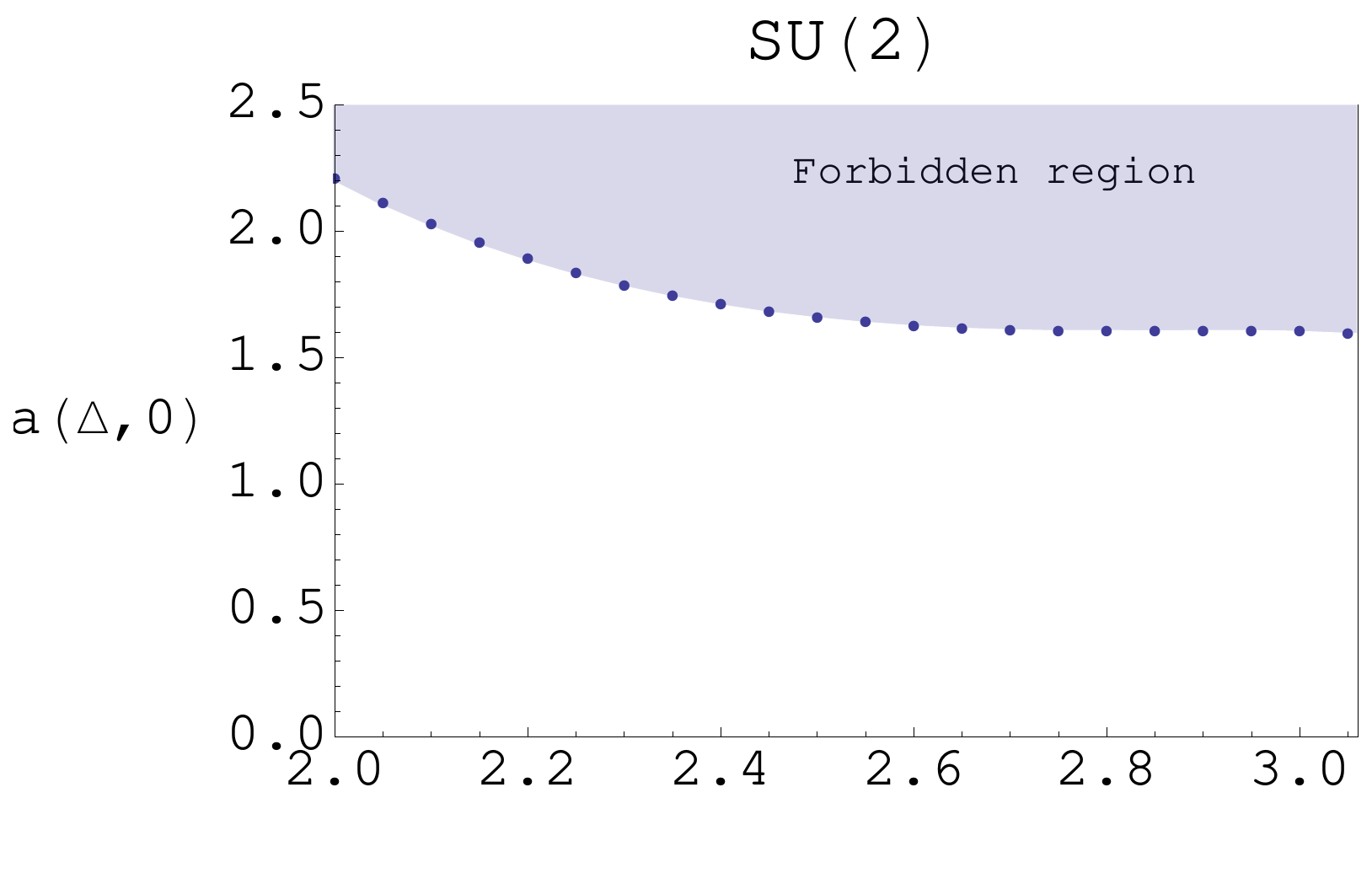}& 
 \includegraphics[width=2.9in]{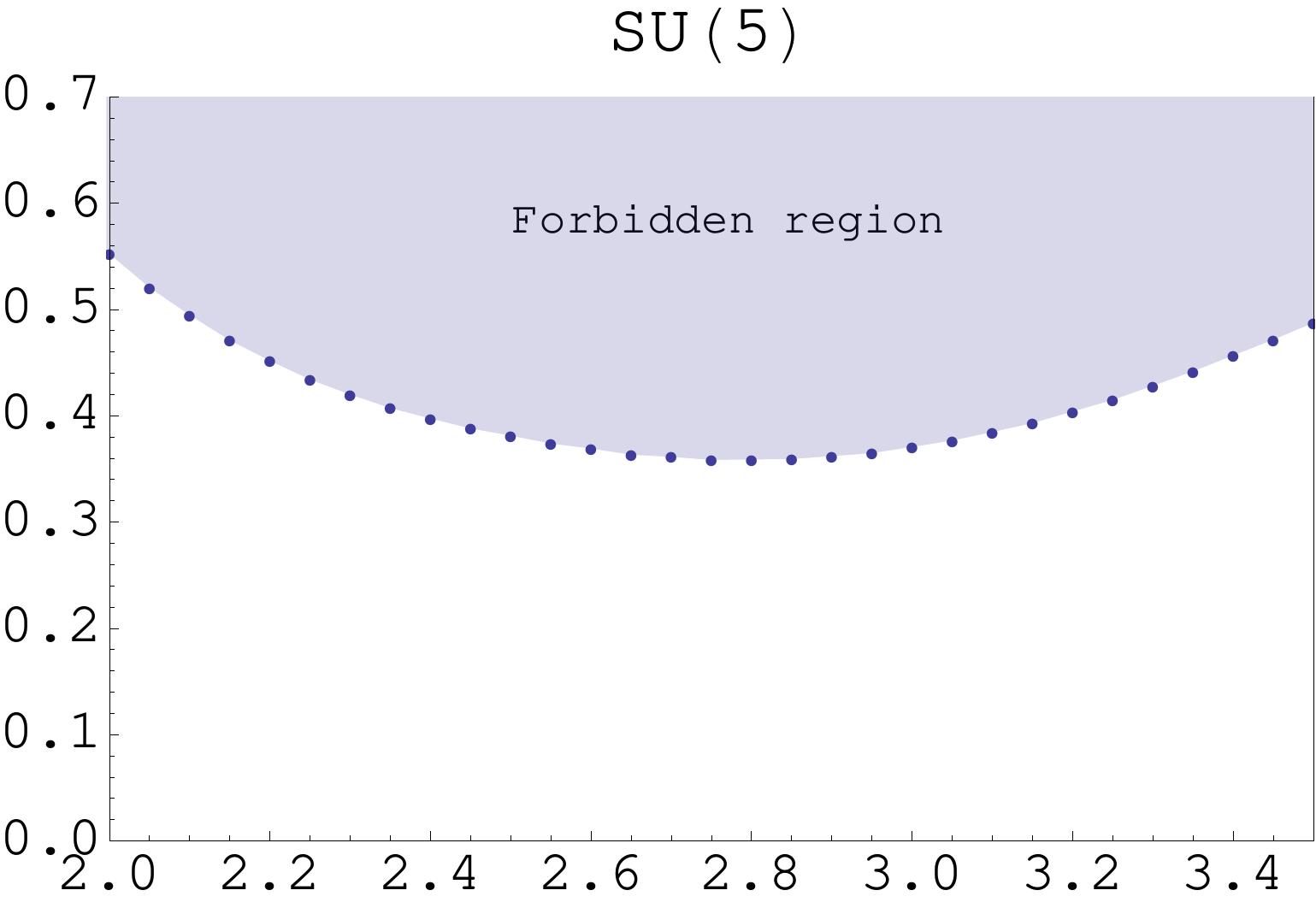} \\
\includegraphics[width=3.25in]{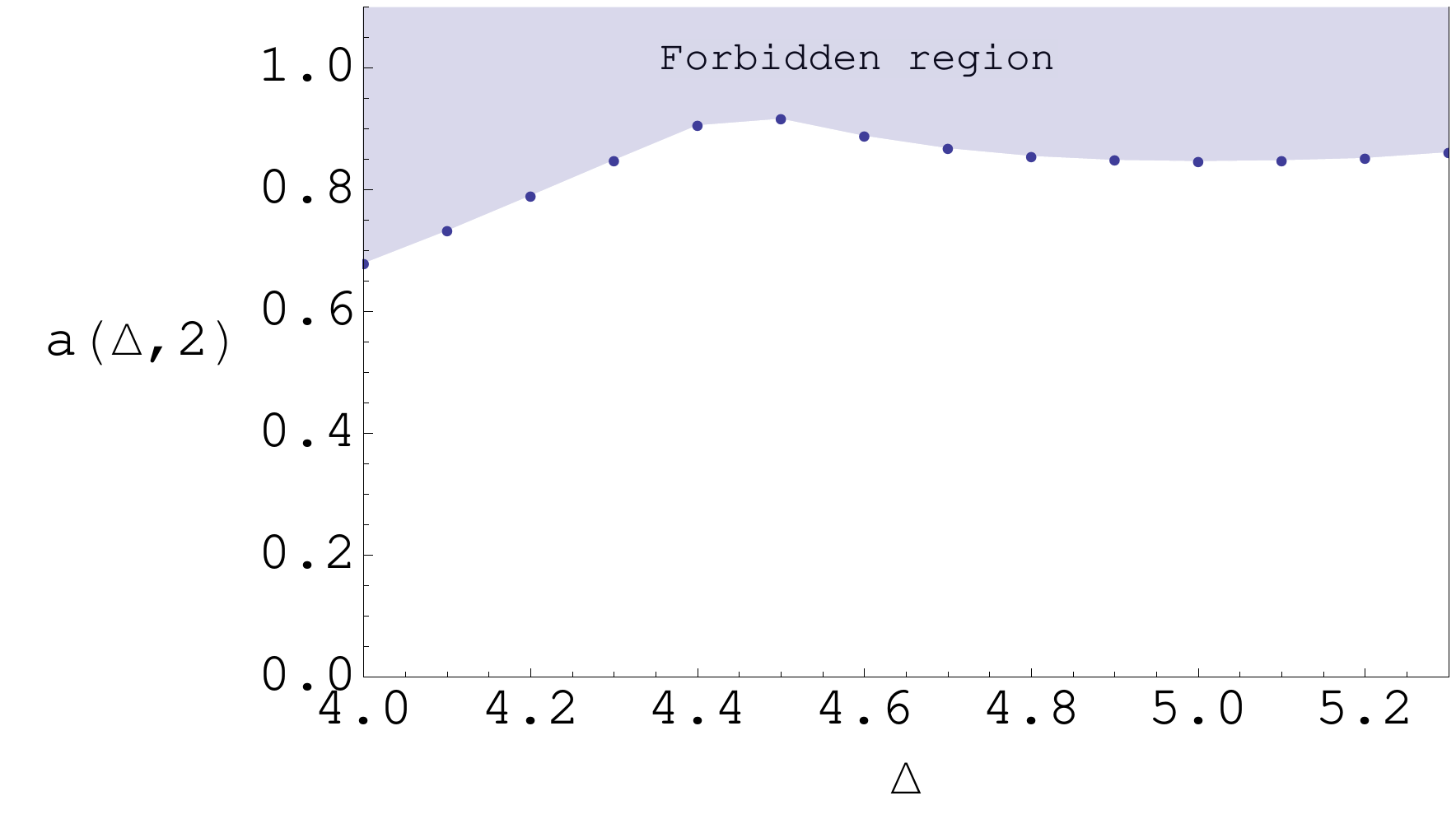} &
\includegraphics[width=2.9in]{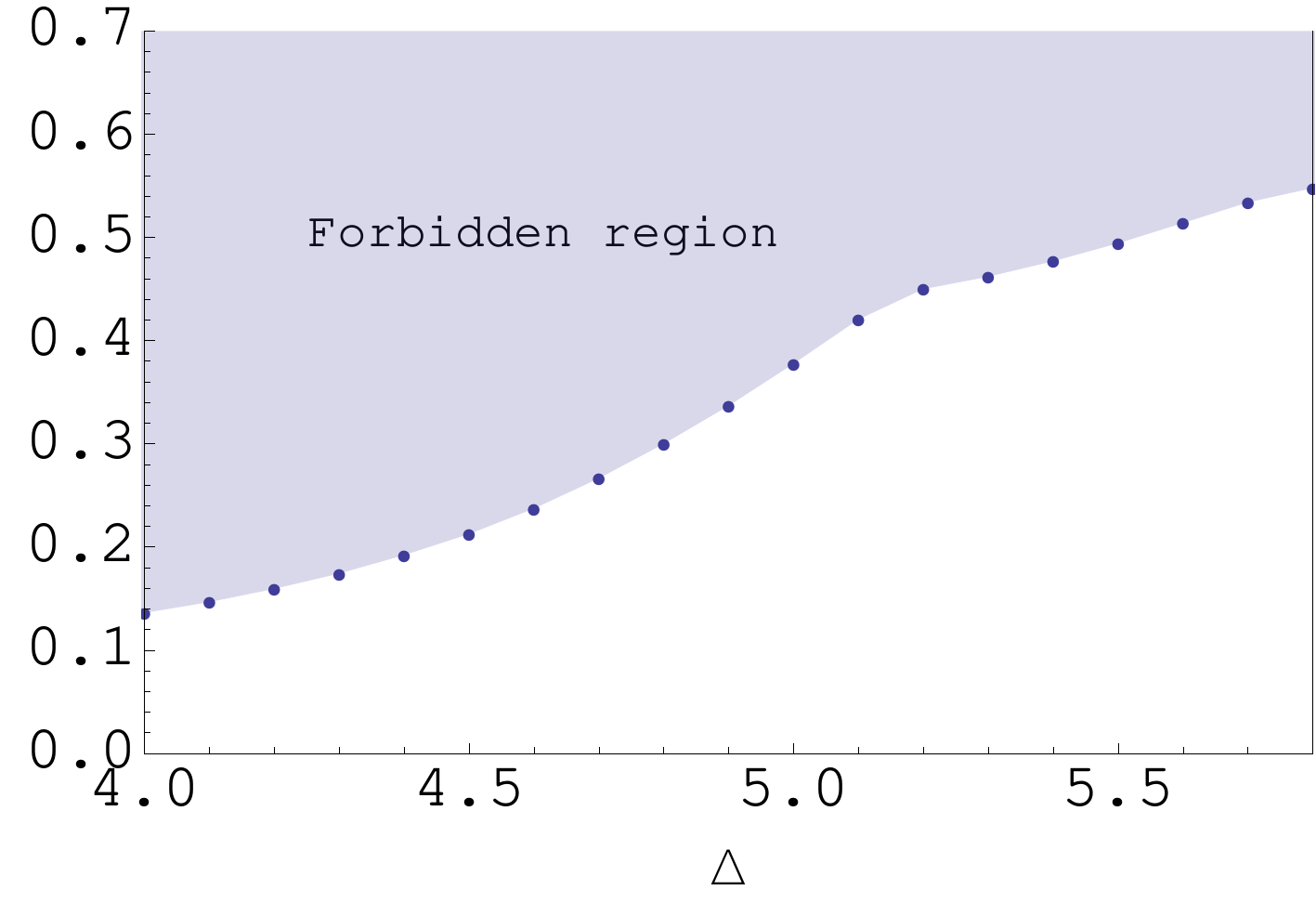}
\end{array}$
\end{center}
\caption{Exclusion regions for the structure constants (or rather, their square) involving two protected operators and one non-protected operator. We show the results for the leading twist operator of spin 0 (top) and spin 2 (bottom), for gauge groups $SU(2)$ (left) and $SU(5)$ (right)}
\end{figure}


Results for other operators can be found in appendix B. These bounds represent robust non-perturbative information about structure constants of ${\cal N}=4$ SYM. In obtaining them we have made no assumptions regarding the spectrum of the theory. Namely, we have considered a continuos spectrum for $\Delta$, consistent with the unitary bound $\Delta \geq \ell+2$. If some information about the spectrum is known, then these constraints can be improved ( in some cases significantly). In the following we will consider two instances of this. 

\section{Particular points of the conformal manifold}

\subsection{Free theory}

Let us analyze the constraints following from the bootstrap equation (\ref{super}) in the free theory limit. In this case, for each fixed $\ell$, we have $\Delta =\ell+2 t$ with $2t$ the twist and $t=1,2,...$. The OPE coefficients have been worked out for this case \cite{Dolan:2001tt}

\begin{eqnarray}
a_{\ell+2,\ell} &=& 2^{\ell+1}\frac{\Gamma^2(\ell+3)}{\Gamma(\ell+5)} \frac{4}{c}\\
a_{\ell+2 t,\ell}&=& 2^\ell \frac{\Gamma(\ell+t+1)\Gamma(\ell+t+2)\Gamma^2(t+1)}{\Gamma(2\ell+2t+1)\Gamma(2t+1)}\left(4(\ell+1)(\ell+2t+2)+\frac{4}{c}(-1)^t \right),~~~~t>1\nonumber
\end{eqnarray}
Nonetheless we can ask the question of finding bounds on the structure constants by assuming the spectrum of the free theory. We have considered $a_{\ell+2t,\ell}$ for several values of the spin and the twist. In all cases by increasing $\Lambda$ the threshold approaches the actual tree-level values for $a_{\Delta,\ell}$, see fig. 2 for two particular examples

\begin{figure}[h]
\begin{center}$
\begin{array}{ccc}
\includegraphics[width=3in]{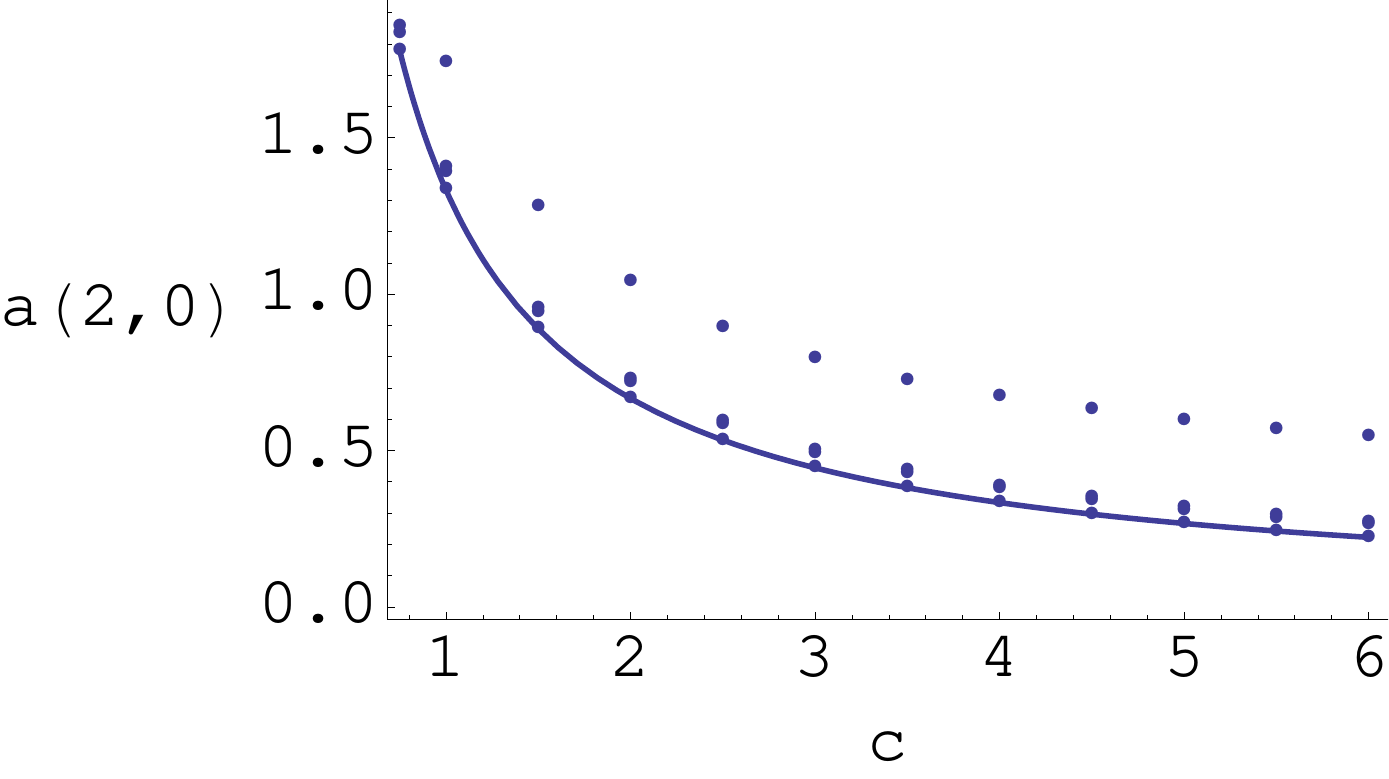}& 
 \includegraphics[width=3in]{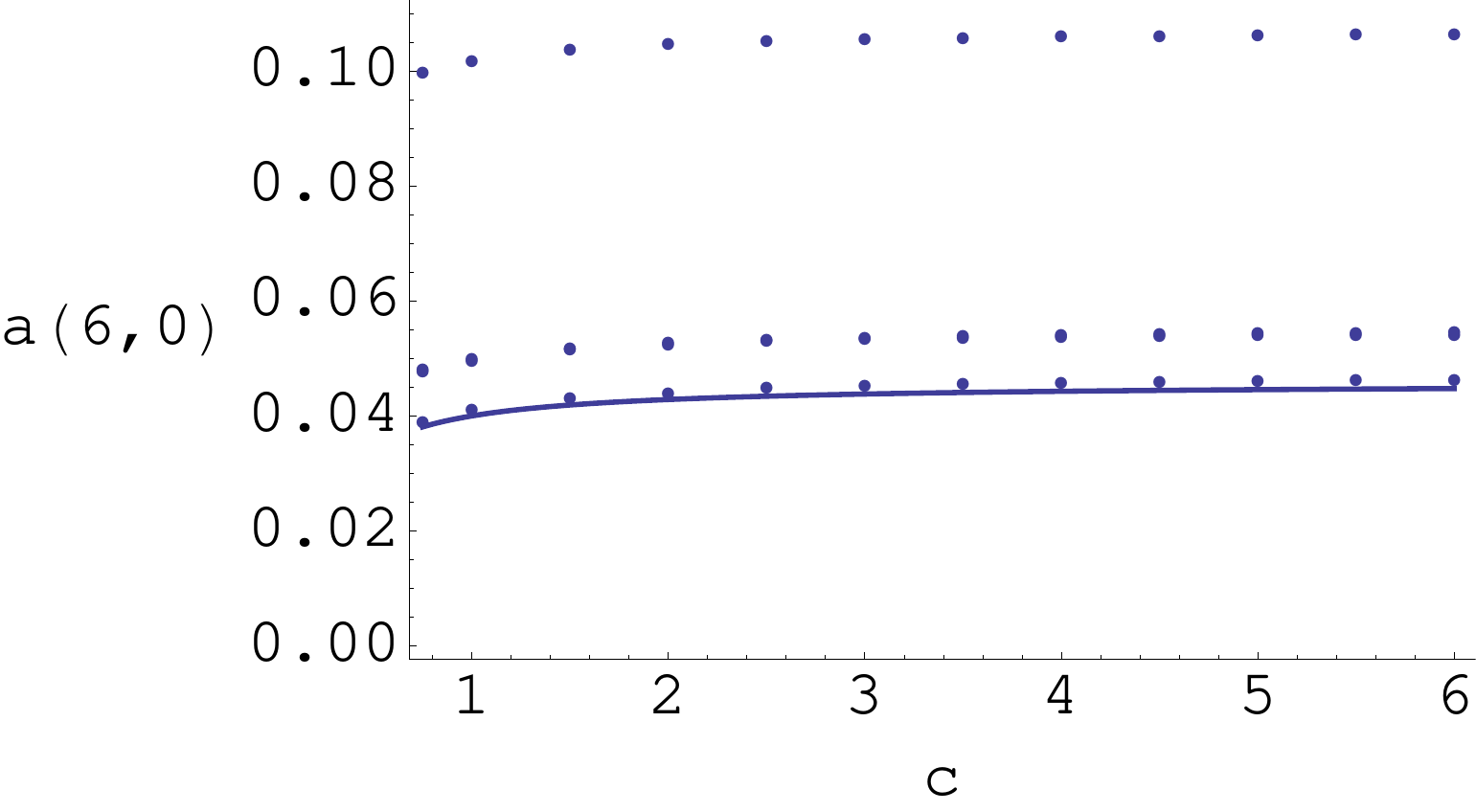} 
\end{array}$
\end{center}
\caption{Threshold values for structure constants as calculated from the conformal bootstrap (assuming the spectrum) vs the actual value (solid line). Different points for a given central charge $c$ denote different values of $\Lambda$. In all cases, as $\Lambda$ increases the threshold values approach the actual value. The figure shows the result for the leading twist operators (left) and twist 6 operator (right) both with zero spin.}
\end{figure}
Maybe this should not come as a surprise: After all for $\ell=0,2,...$ and $\Delta=\ell+2,\ell+4,...$ the conformal blocks form a complete basis of functions and this is the reason why one can solve for the $a_{\Delta,\ell}$. Hence, in the free theory limit, we would expect a unique solution to the superconformal bootstrap equation. In any case, this is a remarkable test of the convergence of the method and shows that the constraints on the OPE coefficients can be improved significantly given information about the spectrum. 

\subsection{Duality invariant points}

Observables of ${\cal N}=4$ SYM are functions of the complexified coupling constant
$$\tau = y +\frac{i}{g}$$
where we have introduced $y= \frac{\theta}{2\pi}$ and $g=\frac{g_{YM}^2}{4\pi}$. ${\cal N}=4$ SYM possesses S-duality under which the coupling constant transforms as
$$ {\bf h} \cdot\tau = \frac{a\tau+b}{c \tau+d}$$
for integers $a,b,c,d$ such that $ad-bc=1$, namely ${\bf h} \in PSL(2,\mathbb{Z})$. $PSL(2,\mathbb{Z})$ contains two finite order subgroups, generated by ${\bf h}_2 \cdot \tau=-1/\tau$ and ${\bf h}_3 \cdot \tau = (\tau-1)/\tau$. Each subgroup has a fixed point in the fundamental domain
$$\tau_2 =i, ~~~~~~\tau_3 = \exp(i\pi/3)$$
In \cite{Beem:2013qxa} it was conjectured that at these values (at least one of them) the bounds on the dimensions derived from the superconformal bootstrap are actually saturated. This implies that at certain value of the coupling constant the spectrum actually satisfies

\begin{eqnarray}
\Delta_0 \geq \Delta_0^{c} \nonumber\\
\Delta_2 \geq \Delta_2^{c} \\
\Delta_4 \geq \Delta_4^{c} \nonumber
\end{eqnarray}
and so on. $\Delta_\ell^{c}$ denotes the threshold (or "corner" value) for the anomalous dimension of the leading twist operator with spin $\ell$. In \cite{Beem:2013hha} the corner values for $\ell=0,2,4$ and for $N=2,3,4$ are provided. We have computed bounds for $a_{\Delta_0^{c},\ell=0}$ and $a_{\Delta_2^{c},\ell=2}$ assuming the above spectrum. A full analysis would require to estimate $\Delta_\ell^{c}$ for $\ell=6,8,...$. Fortunately, the answer depends very weakly on those. For $N=2,3,4$ we find
\begin{eqnarray}
N=2:~~~a_{\Delta_0^{c},\ell=0} < 1.598,~~~~~~ a_{\Delta_2^{c},\ell=2} < 0.805 \nonumber\\
N=3:~~~a_{\Delta_0^{c},\ell=0} < 0.796,~~~~~~ a_{\Delta_2^{c},\ell=2} < 0.586 \\
N=4:~~~a_{\Delta_0^{c},\ell=0} < 0.595,~~~~~~a_{\Delta_2^{c},\ell=2} < 0.547 \nonumber
\end{eqnarray}
Comparing to the previous section, the extra information on the spectrum only improves the bounds slightly for $\ell=0$ but the improvement is significative for $\ell=2$ (for $N=2,3,4$ the values obtained with no assumptions regarding the spectrum are $0.86$, $0.624$ and $0.57$ respectively).

Of course, these bounds could be further improved if we are given extra information about the spectrum of the theory.

In line with the conjectures of \cite{Beem:2013qxa,Beem:2013hha} it is natural to conjecture that these bounds are actually saturated at the duality invariant points. This (if true!) would give the structure constants at a point which no other methods can access. 

\subsection{Interpolating functions and comparison to our results}

In the following we would like to compare our results to appropriate resummations of the available perturbative data. The idea is to construct interpolating functions that give a good approximation to the structure constants to all values of the coupling, and then compare such results with the bounds of the previous section. In order to construct  interpolating functions we will follow  \cite{Beem:2013hha}, who studied the anomalous dimensions of leading twist operators. 

As already mentioned, ${\cal N}=4$ SYM possess S-duality. It was noted in  \cite{Beem:2013hha} that anomalous dimensions of non-protected operators should be invariant under the modular group  $PSL(2,\mathbb{Z})$. We expect the same to be true for the structure constants considered in this paper. \footnote{Namely, involving two-protected scalar operators ${\cal O}_{{\bf 20}'}$ and a superconformal primary in a long multiplet. Complications arise if the protected operators have higher twist and/or descendants are involved. We thank L. Rastelli for discussions on this point.}

The authors of \cite{Beem:2013hha} introduced a refined version of  Pade approximants $P^{(\bf h)}$, manifestly invariants under ${\bf h}$, a finite order subset of the modular transformations.The interpolating functions take the form
\begin{equation}
\label{interpol}
P^{{\bf h}}_{[n/m]}(g) =  \frac{a_0 g_{\bf h}^{(-n)}+ a_1 g_{\bf h}^{(-n+1)}+... +a_n}{g_{\bf h}^{(-m)}+ b_1 g_{\bf h}^{(-m+1)}+...+ b_m}
\end{equation} 
where, for a finite order subgroup of order $d$, the following notation has been introduced
\begin{equation}
g_{\bf h}^{(n)} = \sum_{k=0}^{d-1} ({\bf h}^k.g)^n
\end{equation}
For application to anomalous dimensions, which go as $\gamma \sim g$ at weak coupling, we take $m=n+1$. For application to structure constants, which are finite at tree level, we take $m=n$. Then, the coefficients in (\ref{interpol}) for each observable can be fixed once the perturbative result at $2m$ loops is known. For the generators ${\bf h}_2.\tau=-1/\tau$ and ${\bf h}_3.\tau = (\tau-1)/\tau$ we have
\begin{eqnarray}
g_{{\bf h}_2}^{(n)} = g^n+ \left(\frac{1+y^2 g^2}{g}\right)^n,~~~~~~g_{{\bf h}_3}^{(n)} = g^n+ \left(\frac{1+y^2 g^2}{g}\right)^n+ \left( \frac{1+(1-y)^2 g^2}{g} \right)^n
\end{eqnarray}
We will construct interpolating functions by using the available two-loop results (since four-loop results for structure constants are not available), which are of the form
\begin{eqnarray}
\gamma(g) &=& \alpha_1 g(1+\alpha_2 g^2+... )\\
a(g) &=& \beta_0 \left( 1+ \beta_1 g+\beta_2 g^2+... \right)  
\end{eqnarray}
the interpolation functions in each case read
\begin{eqnarray*}
\label{interpolating}
P^{(2)}_\gamma(g) &=& \alpha_1 \left( \frac{1}{g} +\frac{g}{1+y^2 g^2}-\alpha_2 \right)^{-1},\\
P^{(3)}_\gamma(g) &=& \alpha_1 \left( \frac{1}{g} +\frac{g}{1+y^2 g^2}+\frac{g}{1+(1-y)^2 g^2}-\alpha_2 \right)^{-1}, \\
P^{(2)}_a(g) &=& \beta_0 \frac{\beta_1-\frac{\beta_2}{\beta_1} +\frac{1}{g}+ \frac{g}{1+g^2 y^2}}{-\frac{\beta_2}{\beta_1} +\frac{1}{g} + \frac{g}{1+g^2 y^2}},\\
P^{(3)}_a(g) &=& \beta_0 \frac{\beta_1-\frac{\beta_2}{\beta_1} + \frac{1}{g} +\frac{g}{1+y^2 g^2}+\frac{g}{1+(1-y)^2 g^2}}{-\frac{\beta_2}{\beta_1} + \frac{1}{g} +\frac{g}{1+y^2 g^2}+\frac{g}{1+(1-y)^2 g^2}}
\end{eqnarray*}
One application of these formulas is to get an estimate for the value of the anomalous dimension and structure constant at the fixed point in each case. The fixed points for the order two and three symmetries are $g=1,y=0$ and $g=2/\sqrt{3},y=1/2$ respectively. Let us focus in the case of the spin-zero operator, of the form $Tr \Phi^I \Phi^I$. Using the explicit perturbative results
\begin{eqnarray}
\gamma_0(g) &=& \frac{3N g}{\pi}-\frac{3N^2 g^2}{\pi^2}+...\\
a_0(g) &=& \frac{16}{3(N^2-1)}-\frac{20N}{3(N^2-1)\pi} g+ \frac{N^2}{3(N^2-1)\pi^2}(23+6 \pi^2+72 \zeta_3)g^2+...
\end{eqnarray}
We get the following estimates. These are shown together with the threshold/corner values in tables 1 and 2.
\begin{center}
\begin{table}[htdp]
\hspace{0.5cm}
\begin{minipage}[b]{0.45\linewidth}\centering
\begin{tabular}{|c|c|c|}\hline  & $\gamma_{est.}$ & $\gamma_{th.}$ \\\hline $SU(2)$ & 0.59 - 0.72 & 0.93 \\\hline $SU(3)$ & 0.81 - 0.97 & 1.24 \\\hline $SU(4)$ & 0.99 - 1.17 & 1.47 \\\hline \end{tabular} \caption{Estimated vs predicted values for the anomalous dimension at the duality invariant point}
\end{minipage}
\hspace{0.5cm}
\begin{minipage}[b]{0.45\linewidth}
\begin{tabular}{|c|c|c|}\hline  & $a_{est.}$ & $a_{th.}$ \\\hline $SU(2)$ & 1.59 & 1.598 \\\hline $SU(3)$ & 0.59 & 0.796 \\\hline $SU(4)$ & 0.31 & 0.595 \\\hline \end{tabular}
\caption{Estimated vs predicted values for the structure constant at the duality invariant point}
\label{defaulttable}
\end{minipage}
\end{table}
\end{center}
In all cases, we see that the value given by the interpolating function is of the order of the predicted value. We expect the results to get closer once perturbative results at four loops become available. In any case, note that the agreement for $a_{\Delta_c,0}$ in the case of $SU(2)$ is remarkable. The agreement gets worse as we increase the rank of the gauge group. This is consistent with the expectation that the interpolating functions become less reliable \cite{Beem:2013hha} . In any case, these results show that the conjectured results for $a_{\Delta_c,0}$ at the invariant points are plausible. In particular, for $SU(2)$ we obtain $a_{\Delta_c,0} \sim 1.6$.

We can also make comparisons at generic values of the coupling constant. The interpolating functions (\ref{interpolating}) can be combined in order to obtain $a_{\Delta,\ell}$ as a function of the anomalous dimension $\gamma$. We obtain
\begin{equation}
P_a(\gamma) = \beta_0 \frac{\beta_1-\beta_2/\beta_1+\alpha_2+\frac{\alpha_1}{\gamma}}{-\beta_2/\beta_1+\alpha_2+\frac{\alpha_1}{\gamma}}
\end{equation}
for both ${\bf h}_2$ and ${\bf h}_3$. The following figure shows the behavior of  $P_a(\gamma) $ versus the bounds found in previous section, for the leading twist operators with spin $\ell=0$ and for various gauge groups. 

\begin{figure}[h]
\begin{center}$
\begin{array}{ccc}
\includegraphics[width=2.3in]{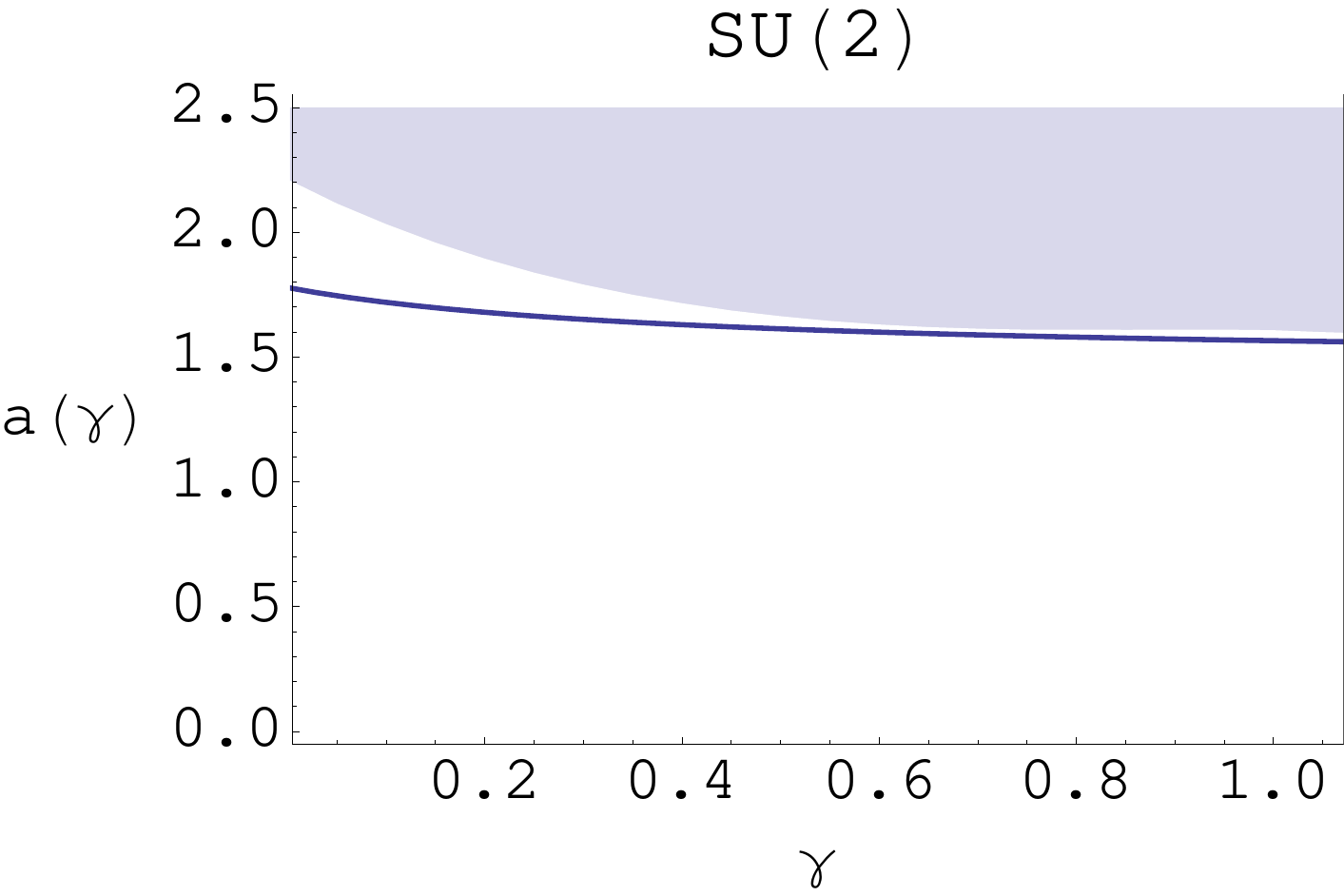}& 
\includegraphics[width=2in]{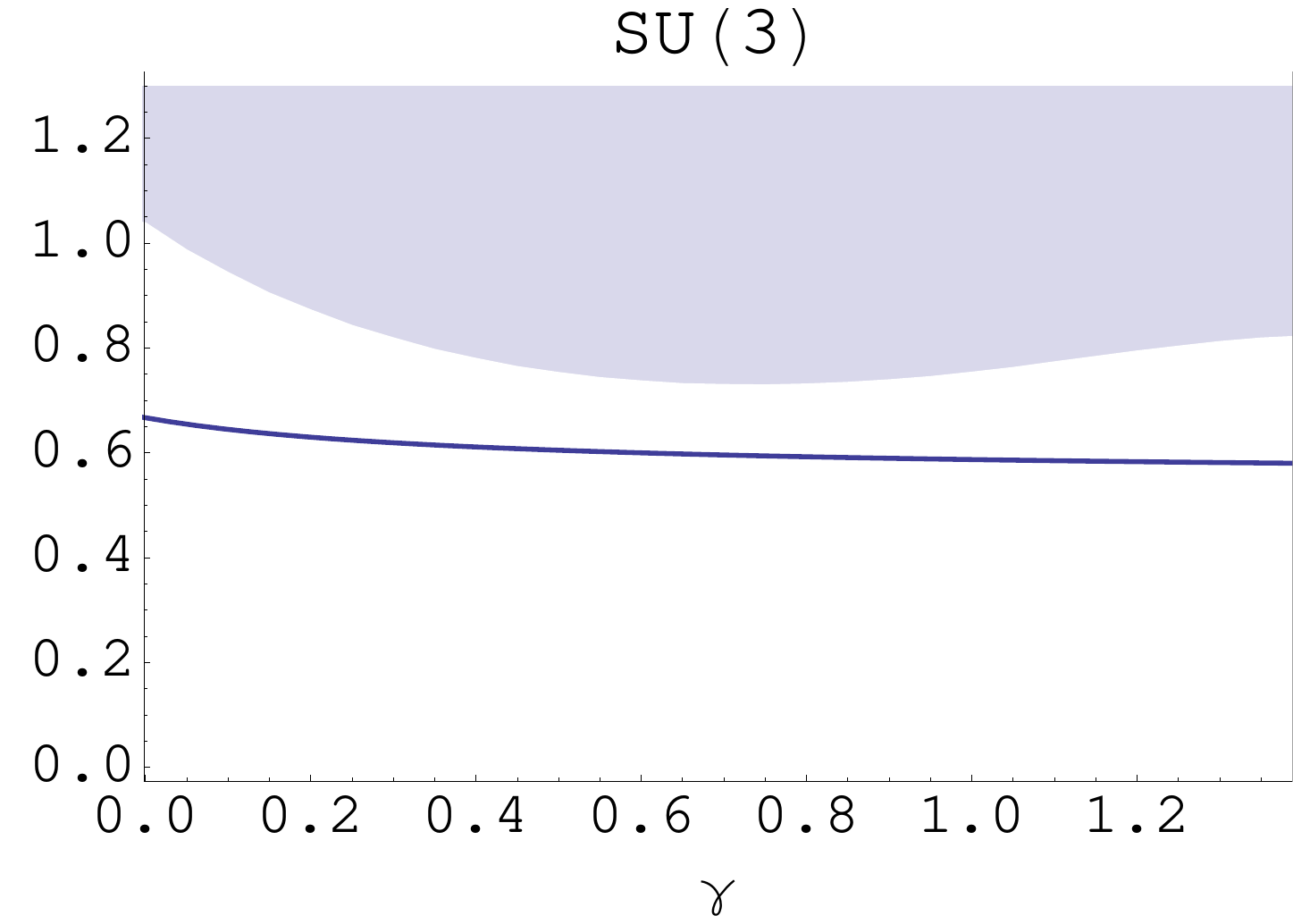} &
 \includegraphics[width=2in]{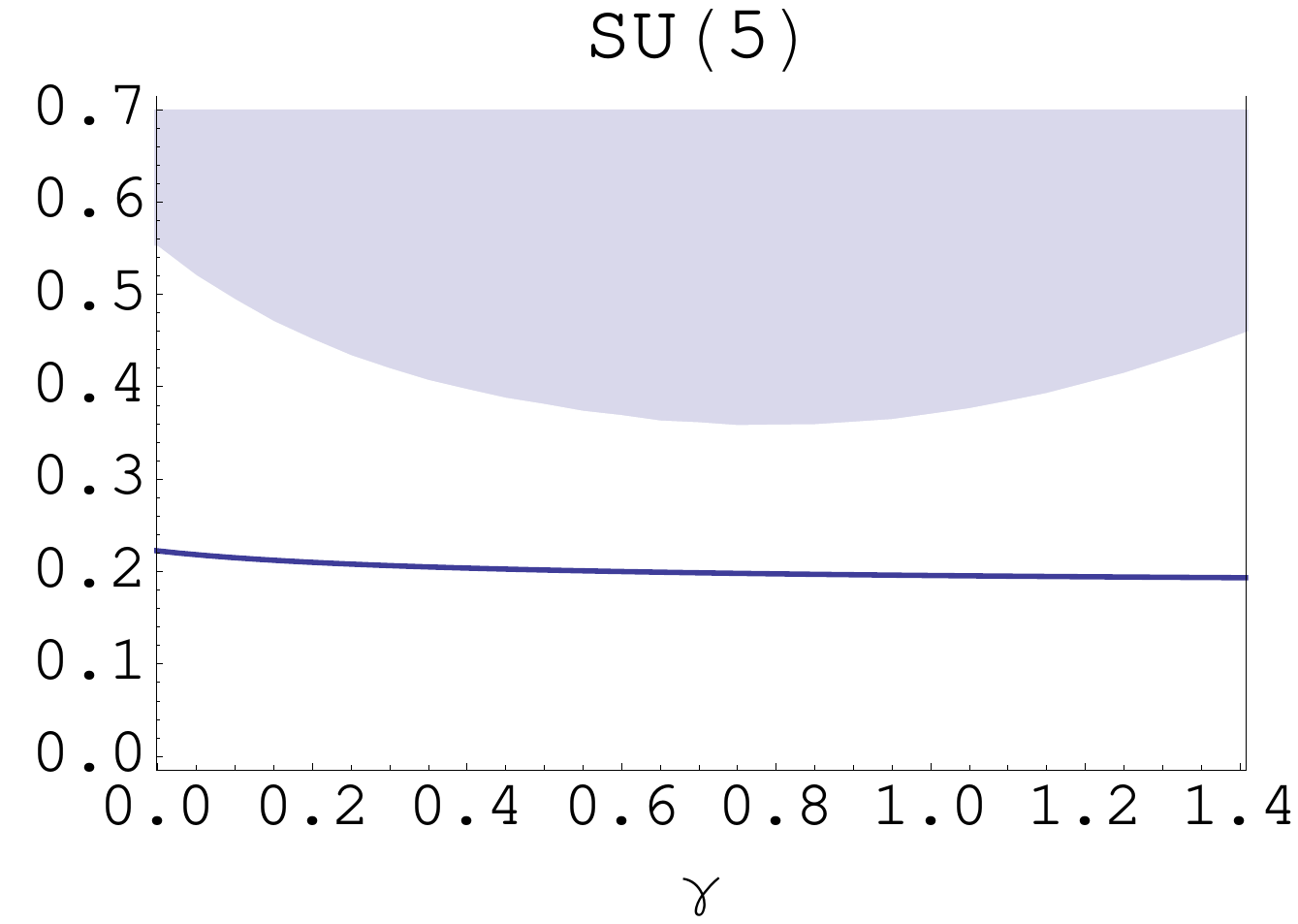} 
\end{array}$
\end{center}
\caption{Interpolating functions vs.  exclusion region, for the structure constant of leading twist operator with $\ell=0$ as a function of the anomalous dimension}
\end{figure}
We observe that all the interpolations are consistent with our bounds, but the bounds can be quite restrictive for small rank gauge groups. We have also considered operators with higher spin, see for instance fig. 4. In this case the bounds appear to be less restrictive
\begin{figure}[h!]
\begin{center}$
\begin{array}{ccc}
\includegraphics[width=2.3in]{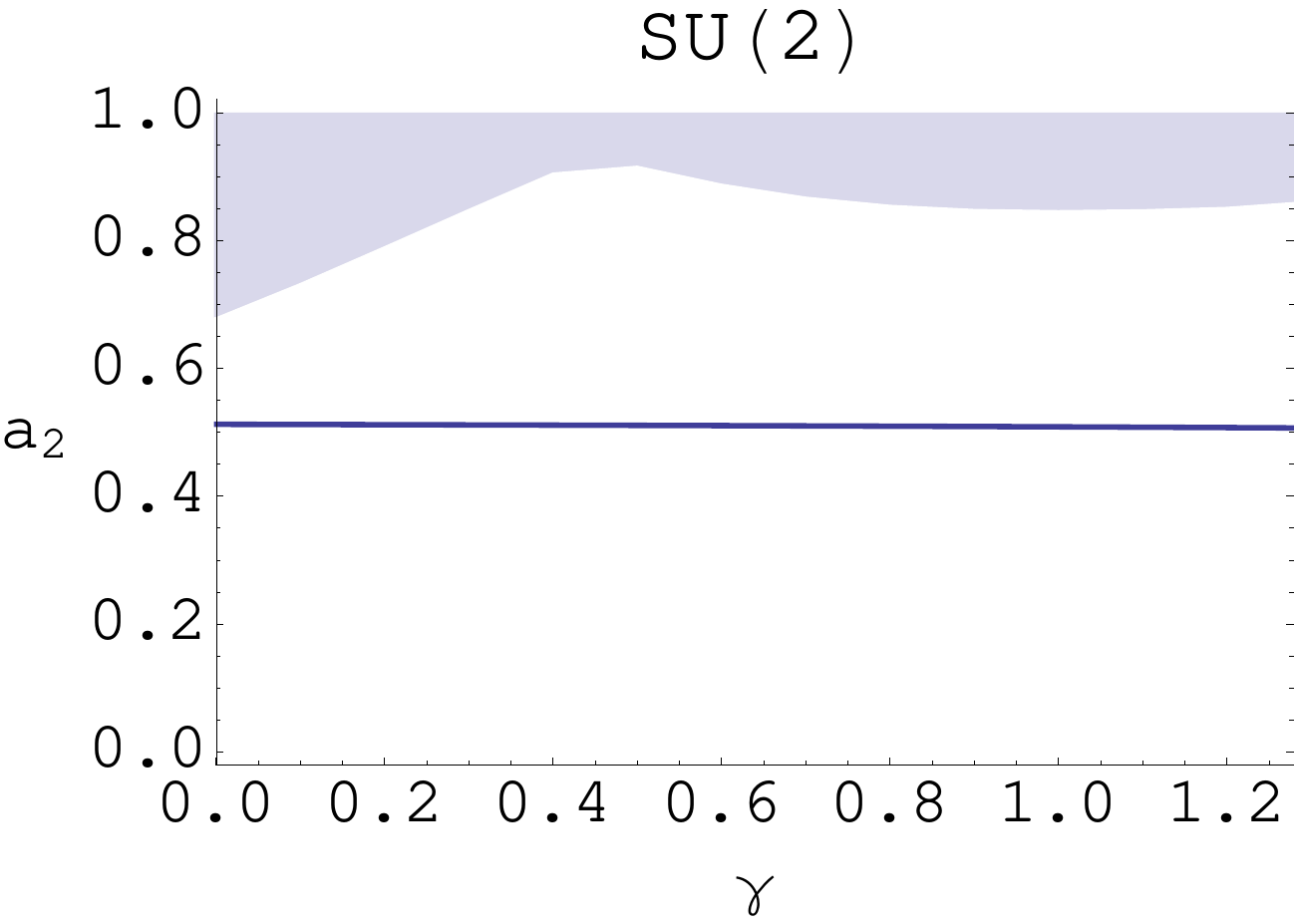}& 
\includegraphics[width=2.3in]{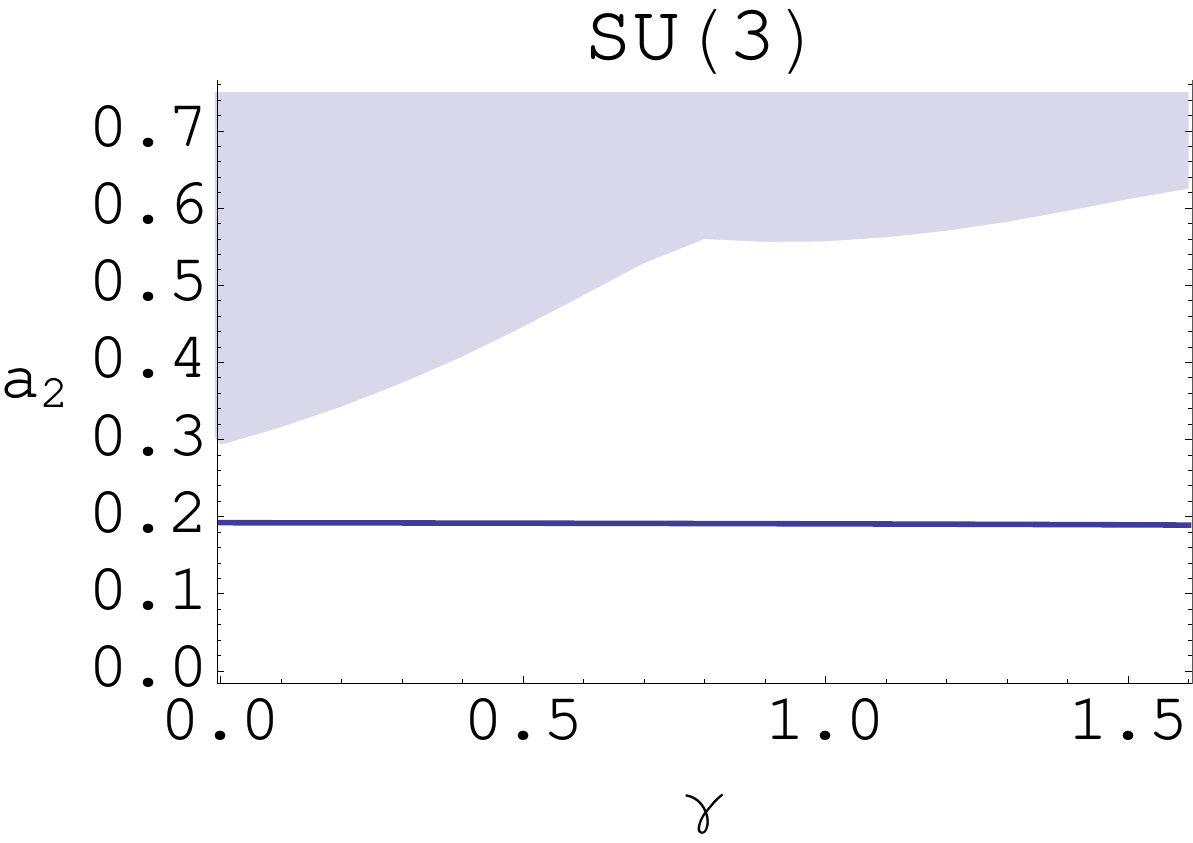} 
\end{array}$
\end{center}
\caption{Interpolating functions vs.  exclusion region, for the structure constant of leading twist operator with $\ell=2$ as a function of the anomalous dimension}
\end{figure}

Finally, note that for $SU(2)$ $\ell=0$, fig. 3,  the interpolating function is very close to the threshold. Interestingly, this is the case were the interpolations are most reliable. This, together with the results of \cite{Beem:2013hha}, suggests that the conformal bootstrap, together with $S-$duality, could be particularly suited to understand ${\cal N}=4$ SYM with gauge group $SU(2)$.

\section{Discussion}

In this letter we have reported on bounds for structure constants/OPE coefficients of ${\cal N}=4$ SYM. These bounds represent exact, non-perturbative, information about  the CFT. 

We have shown that these bounds can be improved at certain points in the conformal manifold, where some information about the spectrum is known. In particular, for the free theory, $g_{YM}=0$, the threshold values given by the conformal bootstrap agree with the known values for the structure constants. Other special points of the conformal manifold are the points fixed under a finite order sub-group of the modular group, for instance $g_{YM}=2\sqrt{\pi}$. In line with conjectures put forward in \cite{Beem:2013qxa,Beem:2013hha} it is natural to propose that at these values of the coupling the structure constants saturate the bounds given by the conformal bootstrap. For instance, for $SU(2)$ gauge group this gives the following prediction
$$a_{\ell=0} \lesssim 1.6$$
for the (square of the) structure constant involving the leading twist operator with spin zero ({\it i.e.} the Konishi operator) at that particular value of the coupling. This is a non-perturbative prediction that would be very hard to make from any other method. If true, this would show the power of the conformal bootstrap (in this case applied to ${\cal N}=4$ SYM) in order to give non-perturbative information about a highly non-trivial CFT.

A method to obtain estimates at arbitrary values of the coupling constant is that of  interpolating functions. In particular one can construct interpolating functions consistent with the symmetries of the theory \cite{Beem:2013hha}. We have compared such estimates to our results and found that they were compatible with the proposal that the bounds are saturated at the duality invariant points. From this comparison one can also see that the constraints seem less stringent for large rank gauge groups. From this point of view the superconformal bootstrap is very different to other approaches to understand ${\cal N}=4$ SYM.

There are several directions that one could follow. In this paper we have used two-loop interpolating functions \footnote{One can only use an even number of loops. This is a limitation of the Pade approximants. This limitation should be circumvented by the interpolating functions constructed in \cite{Sen:2013oza}, but these methods do not apply directly for observables that go as $g^0$ for both, weak and strong coupling.}. At present the structure constants studied in this paper are known to three-loops \cite{Eden:2012rr}. A four-loop result is expected to improve considerably such interpolating functions. By combining the conformal bootstrap with the method of interpolating functions one may get valuable information to any value of the coupling $\tau$, so it would be interesting to pursue this line. 

It would also be interesting to study the dependence of our bounds, and the bounds for the anomalous dimensions, with the spin of the operator. In  \cite{Alday:2013cwa} symmetries of the four-point correlators were used in order to obtain a relation between the structure constants and the anomalous dimensions for leading twist operator of large spin. Such relations are valid to all loops, but still perturbative. It would be interesting to combine this perturbative information with non-perturbative bounds arising from the conformal bootstrap. 

The bounds derived in this paper use very little information about the spectrum of the theory. We have seen that extra information about the spectrum usually results in improved bounds for the structure constants. For instance, it would be interesting to understand the implications of the results of \cite{Komargodski:2012ek} for the problem at hand. 

Finally, it would be fascinating either to prove or disprove the conjectures made in \cite{Beem:2013qxa} and in this paper. Namely, that the bounds for scaling dimensions and structure constants involving leading twist operators are saturated at the duality invariant points.

\subsection*{Acknowledgments}
\noindent
We would like to thank C.~Beem and specially L.~Rastelli for useful discussions. The work of the authors is supported by ERC STG grant 306260. L.F.A. is a Wolfson Royal Society Merit Award holder.

\appendix

\section{Linear operator in the asymptotic region}

In this appendix we consider $F_{\Delta,\ell}$ in the region $\ell,\Delta \gg 1$, with $\ell = \alpha \Delta$. Using the following integral representation for the hypergeometric function
$$~_2F_1(\beta/2,\beta/2,\beta;z)=\frac{\Gamma(\beta)}{\Gamma(\beta/2)^2} \int_0^\infty e^{-\beta/2 t}(1-e^{-t})^{\beta/2-1}(1-z e^{-t})^{-\beta/2}dt$$
together with the method of steepest descent, one can show
\begin{equation}
k_\beta(z) \approx 2^{\beta-1}\frac{(1-z)^{(\beta-1)/4}}{(1+\sqrt{1-z})^{\beta/2-1}(1+\sqrt{1-z}-z)^{\beta/2}},~~~\beta \gg 1
\end{equation}
This allows to write $F_{\Delta,\ell=\alpha \Delta}$ in the asymptotic region in the following form:
\begin{equation}
F_{\Delta,\ell=\alpha \Delta}(z,\bar z)= f(z,\bar z) +f(\bar z,z)-f(1-z,1-\bar z)-f(1-\bar z,1- z),~~~~~f=g h^\Delta
\end{equation}
using variables $a,b$ with $z=1/2+a+b,~\bar z=1/2+a-b$, the small $a,b$ expansion of $g$ and $h$ have the form
\begin{eqnarray}
\label{expansion}
g(a,b) = \frac{g_{0,-1}}{b}+\frac{g_{1,-1} a+ g_{0,0} b}{b}+...\\
h(a,b)= h_{0,0}+h_{1,0}a+h_{0,1}b+...
\end{eqnarray}
Assuming this general expansion, we can act with $\Phi^{(\Lambda)}$ on $F_{\Delta,\ell=\alpha \Delta}(z,\bar z)$. It turns out that only the first leading terms in the expansion (\ref{expansion}) contribute to $\Phi^{(\Lambda)} F_{\Delta,\ell}$ in the large $\Delta$ limit. Furthermore, the leading contribution comes from coefficients $\kappa_{ij}$ such that $2i+2j+1=\Lambda$. We find the result quoted in the body of the paper
\begin{equation}
\Phi^{(\Lambda)} F_{\Delta,\ell=\alpha \Delta} \sim \sum_{2i+2j+1=\Lambda} \frac{(\Lambda+1)!}{(2i)!(2j+2)!} \alpha^{2j+2} \kappa_{ij} 
\end{equation} 
where we have dropped a positive numerical factor independent of $\alpha$ and $\kappa_{ij}$.

\section{Numerical results}

The following shows our numerical results for the bounds on the structure constants. For the tables we have restricted ourselves to $0 \leq \gamma \leq 1$ and a spacing $\delta \gamma =1/10$. These values were obtained with Mathematica and further checked with Matlab and CPLEX optimizer.

\begin{center}
\begin{table}[h!]
\hspace{0.5cm}
\begin{minipage}[b]{0.14\linewidth}\centering
$SU(2),\ell=0$
\begin{tabular}{|c|c|}\hline $\Delta$ & $a_{\Delta,0}^{max}$  \\\hline 2 & 2.208 \\\hline 2.1 & 2.029 \\\hline 2.2 &1.891 \\\hline 2.3 & 1.787 \\\hline 2.4 & 1.712 \\\hline 2.5 & 1.660 \\\hline 2.6 & 1.627 \\\hline 2.7 & 1.610 \\\hline 2.8 & 1.606 \\\hline 2.9 & 1.606 \\\hline 3 & 1.604 \\\hline \end{tabular}
\end{minipage}
\hspace{0.3cm}
\begin{minipage}[b]{0.14\linewidth}
$SU(2),\ell=2$
\begin{tabular}{|c|c|}\hline $\Delta$ & $a_{\Delta,2}^{max}$  \\\hline 4 & 0.678 \\\hline 4.1 & 0.732 \\\hline 4.2 & 0.789 \\\hline 4.3 & 0.848 \\\hline 4.4 & 0.905 \\\hline 4.5 & 0.915 \\\hline 4.6 & 0.887 \\\hline 4.7 & 0.867 \\\hline 4.8 & 0.854 \\\hline 4.9 & 0.848 \\\hline 5 & 0.846 \\\hline \end{tabular}
\end{minipage}
\hspace{0.2cm}
\begin{minipage}[b]{0.14\linewidth}
$SU(3),\ell=0$
\begin{tabular}{|c|c|}\hline $\Delta$ & $a_{\Delta}^{max}$  \\\hline 2 & 1.038 \\\hline 2.1 & 0.944 \\\hline 2.2 & 0.872 \\\hline 2.3 & 0.819 \\\hline 2.4 & 0.780 \\\hline 2.5 & 0.753 \\\hline 2.6 & 0.736 \\\hline 2.7 & 0.730 \\\hline 2.8 & 0.731 \\\hline 2.9 & 0.739 \\\hline 3 & 0.753 \\\hline 
 \end{tabular}
\end{minipage}
\hspace{0.2cm}
\begin{minipage}[b]{0.14\linewidth}
$SU(3),\ell=2$
\begin{tabular}{|c|c|}\hline $\Delta$ & $a_{\Delta}^{max}$  \\\hline 4 & 0.292 \\\hline 4.1 & 0.315 \\\hline 4.2 & 0.342 \\\hline 4.3 & 0.372 \\\hline 4.4 & 0.406 \\\hline 4.5 & 0.445 \\\hline 4.6 & 0.486 \\\hline 4.7 & 0.527 \\\hline 4.8 & 0.559 \\\hline 4.9 & 0.555 \\\hline 5 & 0.556 \\\hline 
 \end{tabular}
\end{minipage}
\hspace{0.2cm}
\begin{minipage}[b]{0.14\linewidth}
$SU(5),\ell=0$
\begin{tabular}{|c|c|}\hline $\Delta$ & $a_{\Delta}^{max}$  \\\hline 2 & 0.551 \\\hline 2.1 & 0.494 \\\hline 2.2 & 0.451 \\\hline 2.3 & 0.419 \\\hline 2.4 & 0.397 \\\hline 2.5 & 0.381 \\\hline 2.6 & 0.368 \\\hline 2.7 & 0.361 \\\hline 2.8 & 0.359 \\\hline 2.9 & 0.361 \\\hline 3 & 0.370 \\\hline 
 \end{tabular}
\end{minipage}
\hspace{0.2cm}
\begin{minipage}[b]{0.14\linewidth}
$SU(5),\ell=2$
\begin{tabular}{|c|c|}\hline $\Delta$ & $a_{\Delta}^{max}$  \\\hline 4 & 0.135 \\\hline 4.1 & 0.146 \\\hline 4.2 & 0.159 \\\hline 4.3 & 0.173 \\\hline 4.4 & 0.191 \\\hline 4.5 & 0.212 \\\hline 4.6 & 0.236 \\\hline 4.7 & 0.266 \\\hline 4.8 & 0.299 \\\hline 4.9 & 0.336 \\\hline 5 & 0.377 \\\hline 
 \end{tabular}
\end{minipage}
\caption{Numerical bounds for structure constants of leading operators of spin $\ell=0,2$ and dimension $\Delta$. In each case, it is assumed that the operator exists. This should be the case for $\Delta$ smaller than the "corner" values found in \cite{Beem:2013hha}.}
\end{table}
\end{center}


\end{document}